# Influence of microstructure on superconductivity in $K_xFe_{2-y}Se_2$ and evidence for a new parent phase $K_2Fe_7Se_8$

Xiaxin Ding[1], Delong Fang[1], Zhenyu Wang[2], Huan Yang[1], Jianzhong Liu[1], Qiang Deng[1], Guobin Ma[1], Chong Meng[1], Yuhui Hu[1] & Hai-Hu Wen[1]

[1]National Laboratory of Solid State Microstructures and Department of Physics, Nanjing University, Nanjing 210093, China. [2]National Laboratory for Superconductivity, Institute of Physics and National Laboratory for Condensed Matter Physics, Chinese Academy of Sciences, Beijing 100190, China.

Correspondence and requests for materials should be addressed to H.-H.W. (email: hhwen@nju.edu.cn).

The search for new superconducting materials has been spurred on by the discovery of iron-based superconductors whose structure and composition is qualitatively different from the cuprates. The study of one such material, $K_xFe_{2-y}Se_2$ with a critical temperature of 32 K, is made more difficult by the fact that it separates into two phases - a dominant antiferromagnetic insulating phase with a $K_2Fe_4Se_5$ structure, and a minority superconducting phase whose precise structure is as yet unclear. Here we perform electrical and magnetization measurements, scanning electron microscopy and microanalysis, X-ray diffraction, and scanning tunnelling microscopy on $K_xFe_{2-y}Se_2$ crystals prepared under different quenching processes to better understand the relationship between its microstructure and its superconducting phase. We identify a 3D network of superconducting filaments within this material and present evidence to suggest that the superconducting phase consists of a single Fe vacancy for every eight Fe-sites arranged in a $\sqrt{8}\times\sqrt{10}$ parallelogram structure.

## Introduction

Since the discovery of high temperature superconductivity in F-doped LaFeAsO (ref.1), many new iron based superconductors with different structures have been fabricated[2]. A new superconducting system, namely, the $K_xFe_{2-y}Se_2$, was discovered by the end of 2010 (ref.3, 4). This material has the iso-structure of the FeAs-based 122 superconductors[5,6,7] and brings about new interests and curiosity to the community because the preliminary band structure calculation on this system[8] has predicted the absence of the hole pocket which is supposed to be necessary for the theoretical picture of S± pairing[9,10]. Soon later, it was suggested that the material might be iron deficient and separate into the mixture of an insulating antiferromagnetic $K_2Fe_4Se_5$ phase and a superconducting phase[11,12,13]. Up to now the phase separation in superconducting $K_xFe_{2-y}Se_2$ single crystals has been intensively investigated and widely perceived[14].

From the fact that a robust superconducting magnetic shielding signal appears at low magnetic fields, but it becomes much weaker at high magnetic fields, we point out the percolation of the superconducting state in $K_xFe_{2-y}Se_2$ (ref.11). This picture can easily explain the double magnetic penetrations in the low magnetic field region on the magnetic-hysteresis-loops (MHL) and the very narrow MHL width[11] which is proportional to the superconducting critical current density according to the Bean critical state model. Later, nanoscale phase separation was also inferred from the scanning tunneling microscopy (STM)[12] and angle resolved photoemission spectroscopy (ARPES)[13]. Earlier measurements on Mössbauer [15] effect in the same sample reveal a magnetic phase (later shown to be the antiferromagnetic insulating phase $K_2Fe_4As_5$)[16,17] which occupies about 80-90% of the volume, leaving only 10-20% of the volume as the paramagnetic phase which becomes superconductive at low temperatures. Phase separation has also been found in $K_xFe_{2-y}Se_2$ by many other techniques[18,19].

By quenching the original $K_xFe_{2-y}Se_2$ single crystals from a high annealing temperature (e.g., 350 °C), we found that the global appearance of superconductivity inferred from the resistive and the magnetic measurements was significantly improved[20]. It is curious to know why the quenching process affects superconductivity so obviously. One proposal would be that, after being quenched, the iron vacancies may be in the disordered state which somehow suppresses the antiferromagnetic state and promotes the superconductivity[21]. The phase separation was investigated by Liu *et al.*[22] and Speller *et al.*[23] with a great detail by analyzing the correlation between the microstructures and superconductivity. Meanwhile, nanoscale phase separation has also been suggested from the scanning nano-focused x-ray diffraction[24]. It remains unresolved whether the enhanced superconductivity in the quenched samples is induced by the disorder of the iron vacancies, or the superconductivity appears in an ordered state with an unknown microstructure of the iron vacancies. Furthermore, it is curious to know what is the parent phase from which the superconductivity is derived. Here we demonstrate that the superconductivity has a close relationship with the microstructures due to the phase separation, and for the first time, we propose that the superconducting filamentary paths are derived from a possible parent phase with one vacancy in every eight Fe sites, forming a unique $\sqrt{8}\times\sqrt{10}$ parallelogram structure.

## Results

**Sample preparation and characterizations.** The original $K_xFe_{2-y}Se_2$ single crystals were fabricated by using the self-flux method. Details for growing the samples are presented in Methods. We prepare the target samples with different microstructures by following the quenching technique established early in our group[20]. The original single crystals have been thermally treated in three different ways: slow furnace cooled (samples named as SFC) from the high sintering temperature (1030 °C), retreated for 2 hours at 250 °C (samples named as S250) and 350 °C (samples named as S350) and followed by quenching in liquid nitrogen. In Figure 1a, we show the temperature dependence of resistivity of the three typical samples. It is clear that the slow-furnace-cooled sample SFC exhibits a broad superconducting transition which starts at about 25 K, which is accompanied by a large residual resistivity (~ 450 mΩ-cm). For the quenched sample S350, as shown in Figure 1b, a sharp superconducting transition occurs at about 32 K with a quite small residual resistivity (8.4 mΩ-cm), indicating a much better connectivity of the superconducting phases. This evolution of superconductivity with quenching is consistent with the magnetization data. As shown in Figure 1c, the magnetization measured at 20 Oe using the zero-field-cooled (ZFC) mode on the sample SFC is very small, indicating a poor connectivity of superconducting phases. This is in sharp contrast with the sample of S350. A simple estimate on the magnetization of S350 measured in the zero-field-cooled mode at $H$ = 20 Oe gives a magnetic shielding volume of about 80%. However, as we addressed previously[11], for a superconductor constructed by filamentary networks, it is not valid to calculate the superconducting volume using the magnetic shielding signal measured at a low magnetic field. In Figure 2a, the magnetization-hysteresis-loops (MHLs) of the three typical samples are presented. The SFC sample exhibits a very tiny irreversible magnetization, indicating a rapid flux motion. While for the sample S350, the MHL becomes much wider, indicating a certain strength of flux pinning. In Figure 2b, the MHLs measured at different temperatures from 2 K to 28 K are presented. It is interesting to see that, even up to 28 K (close to $T_c$), the MHL is still open showing a capability of carrying superconducting current up to about 2 T. This suggests that the superconducting area, although being small in volume fraction, has robust superconductivity as in the FeAs-based 122 samples. The behavior of magnetization in $K_xFe_{2-y}Se_2$ samples after different thermal treatments was also observed by other groups[22,25]. Here we use the Bean critical state model, although it may be inappropriate for an inhomogeneous superconductor, to calculate the superconducting critical current density ($J_c$) and present the data in Figure 2c. At 2K and near zero field, we found a value of $J_c$=1.4 ×10$^4$ A/cm$^2$, which is about 1 to 2 orders of magnitude lower than that of the FeAs-based 122 samples[26]. The much suppressed $J_c$ and the easy magnetic flux penetration in $K_xFe_{2-y}Se_2$ are consistent with the picture of percolative superconductivity. Comparing the transport and the magnetic properties among the three typical samples, we conclude that the phase separation should be more severe in SFC than in S350.

**SEM measurements.** In order to illustrate the relationship between the microstructures and superconductivity in three different samples, we measured the morphology on the surface of newly cleaved samples using an advanced field-emission scanning electron microscope (SEM). The details for the SEM measurements are given in Methods. For three typical samples, as shown in Figure 3, they all separate into two major regions, which are characterized by the domain-like brighter area and a background with darker color. A close scrutiny finds that there are some differences among the microstructures of the three typical samples. For the sample SFC (shown in Figure 3a and 3b), the brighter domains, with roughly rectangular shapes, have a larger size and are well separated from each other. For the sample S250 (shown in Figure 3c and 3d), the large domains in SFC further crack but are still well separated. As shown in Figure 3e and 3f, for the sample S350, however, the brighter area split into many tiny domains which uniformly spread out and closely connect into a spider-web-like network. Similar features were observed by other groups[22,23]. In order to get a deeper insight on the two different areas exposed by our SEM, we did the measurements of local composition across certain areas in the samples. As shown in Figure 4a for the sample of SFC, we cleaved the sample and randomly select an area with these two typical kinds of regions and some naturally formed terraces. Then we randomly select two brighter domains and collect the data of local compositions of K, Fe and Se by scanning along the trace highlighted by the yellow arrowed line. It is clear that the Fe (K) concentration is higher (lower) on the domains than in the background. As presented in Figure 4b, a close correlation between the local compositions of K and Fe is observed. By figuring out the local compositions, we find that the brighter domains have a composition of about $K_{0.64}Fe_{1.78}Se_2$, while the background area has a composition of about $K_{0.81}Fe_{1.60}Se_2$, the latter is very close to the standard $K_2Fe_4Se_5$ phase[16,17]. To strengthen these conclusions, we did the local analysis on 50 randomly selected specific points, marked by the red spots (on background) and blue spots (on the domains) in Figure 4c. The statistics on these data are given in Figure 4d. Interestingly the data are rather converged and fall into mainly two groups, one with the composition of about $K_{0.68}Fe_{1.78}Se_2$ and another one around $K_{0.8}Fe_{1.63}Se_2$. In combination with the transport and magnetic properties mentioned above, and taking the composition analysis into account, the brighter areas are naturally attributed to the superconducting filamentary paths, while the background corresponds to the well understood antiferromagnetic 245 phase. Further analysis on the sample S350 gives the similar compositions in two different regions as the SFC.

To know whether the fractional area occupied by the domains will change after different thermal treatments, we estimate the fractional ratio of the domains (suppose to be the superconducting areas) of the three typical samples. Since the SEM images are all gray-scale digital images, we set a gray value, and then converted the color of the domains to white and the color of the background to black. Then we let the program to calculate the fractional ratio of white areas over the total. It is interesting to note that, the fractional ratio of the area occupied by the brighter domains in all three kinds of samples are close to each other: SFC (20.51%), S250 (18.23%), S350 (16.97%). Clearly the fractional ratio occupied by the superconducting areas is quite stable, although a trend of slight decreasing after the quenching has been observed. Based on above phase separation scenario, we conclude that the quenching procedure itself doesn't create superconducting area, however it changes the arrangement of superconducting areas and probably also changes slightly the chemical phase in the quenched samples.

Another interesting observation here is that, these superconducting domains exist actually as three dimensional pillars buried in the 245 background. This can be evidenced at several places marked by the circles in Figure 4a, where the terraces exist due to the cleaving, but the superconducting domains extend in many layers along the c-axis. In this sense, the superconducting filamentary paths form a three dimensional network, which we call as the 3D spider-web-like network. Similar behavior of the composition distribution is observed in the sample S350 which presents better global appearance of superconductivity, although now the domains become much smaller. Therefore our data clearly illustrate the 3D spider-web-like superconducting filamentary paths.

The phase separation can also be concluded from the x-ray diffraction pattern (XRD) measurements performed on a series of single crystal samples. Figure 5a shows the XRD data of the slow furnace cooled and quenched crystals. The sharp diffraction peaks indicate excellent crystalline quality of all crystals, and the peak splitting of the (00*l*) reflections evidences the coexistence of two spatially separated phases in the crystals. As can be seen, there is hardly any obvious shifting among the peaks of the majority phase which corresponds to the AFM ordered 245 structure. The peaks from the minority phase which corresponds to the domains in the SEM image are marked by the asterisks. From the enlarged view of (008) and $(00\underline{10})$ reflections (Figure 5b), we can see that the peaks indicated by the asterisks of S250 stand still, at the same positions of the SFC. In contrast to S250, the peaks from the minority phase of S350 seem to shift slightly and become closer to the majority phase. Taking a comparison with the SEM images, the results suggest that the chemical phase of the superconducting areas doesn't change and the "islands" just crack into smaller pieces when quenched at 250 ℃. Instead, the "domains" change into many thin but well connected networks and the c-axis shrinks slightly in the S350 samples.

**Discussion**

In our local analysis presented above, it is found that the composition of the superconducting domains are converged at about $K_{0.68}Fe_{1.78}Se_2$. Assuming that the Fe and Se atoms on the domains have a basic valence state of $Fe^{2+}$ and $Se^{2-}$ respectively, the average doping level is found to be around 0.12 e/Fe. This value is actually very close to that determined by the Fermi surface area in ARPES which gives 0.11 e/Fe (ref.27) in the same sample. Concerning the uncertainties in these totally distinct techniques, the consistency between the two techniques are remarkable, which also validates the argument that the domains are superconducting and have a composition of about $K_{0.68}Fe_{1.78}Se_2$. In the background phase the analysis show a composition of $K_{0.8}Fe_{1.63}Se_2$, which is naturally attributed to the well-known 245 phase. This further validates the composition analysis on the superconducting domains. Concerning what is the real superconducting phase and/or the parent phase for superconductivity, it remains to be highly controversial. The local STM measurements of topography and tunneling spectrum on $KFe_2Se_2$ thin films suggest that the superconductivity may arise from the phase with the standard formula $KFe_2Se_2$ or slightly Se-deficient $KFe_2Se_{2-x}$ (ref.12). Recently the same group has concluded that the Fe-vacancy free FeSe planes are superconductive when they are in proximity to the 245 phase[28]. In sharp contrast, Texier et al.[29] recently concluded indirectly from the NMR measurements that the superconducting phase might have the formula $Rb_{0.3}Fe_2Se_2$. From the point of view of the local stress induced by the electrostatic field, the formula of $KFe_2Se_2$ is actually in an extreme case where the system needs to overcome a huge electric potential, which may be only realized in thin films. Meanwhile the structure with the formula of $Rb_{0.3}Fe_2Se_2$ may be also in the verge of instability because the K content is too little to sustain the 122 structure. Therefore the two confronting issues in the system of $K_xFe_{2-y}Se_2$ are that, on one hand one needs to lower down the electrostatic force by reducing both the Fe concentration from 2 and the K concentration from 1, and on the other hand it is necessary to sustain the 122 structure. The 245 phase is a naturally balanced state between these two limits: the charges are well balanced by having the ionic states of $K^+$, $Fe^{2+}$ and $Se^{2-}$, and the Fe and K contents are high enough to sustain the 122 structure. The formation of the long range antiferromagnetic order can further lower down the energy of the system. This is also the reason that the 245 phase is very easy to be formed in the fabrication. A glance at our formula $K_{0.68}Fe_{1.78}Se_2$ of the superconducting domains, would suggest that a possible phase with the formula of $K_{0.5}Fe_{1.75}Se_2$ (or call as 278 phase) is next to the $K_{0.8}Fe_{1.6}Se_2$ (or the 245) phase and may play as the parent phase for superconductivity. This parent phase has one vacancy in every eight Fe-sites. A slight adding of K atoms and Fe atoms in $K_{0.5}Fe_{1.75}Se_2$, as usually happening in the synthesizing process, will lead to the electron doping and induce superconductivity. As far as we know, no any theoretical or experimental results about this 278 phase have been reported.

In order to check the proposed formula or structure of the Fe-lattice, i.e., one Fe vacancy in every eight Fe-sites (or named as 1/8 Fe-vacancy state), we have tried to use STM to search for the possible evidence. After cleaving the single crystals of $K_xFe_{2-y}Se_2$, the most possible top layers exposed for the STM measurements are constructed by K or Se atoms. Thus it is very difficult to resolve the structure of the Fe-layer in the cleaved single crystals. However, when the K-atom layer is exposed as the top layer, according to the basic understanding of local charge balance, a possible case is that on top of each Fe-vacancy there exists one K atom. This, as explained schematically in Figure 6c and 6d, can significantly release the local stress induced by the electrostatic force. Thus it is quite possible to observe a layer of K atoms, which has a lattice with the same structure of the Fe-vacancy in the beneath layer. Actually, for the case of one Fe-vacancy out of eight Fe-sites, there are at least five kinds of spatial arrangement for the Fe vacancies, for example $2\times4$, $2\times\sqrt{17}$, $\sqrt{5}\times\sqrt{13}$, $\sqrt{8}\times\sqrt{8}$ and $\sqrt{8}\times\sqrt{10}$ (in units of Fe-Fe bonds) with different block shapes. By doing the STM measurements on the domains, we sometimes observe a lattice with the structure as shown in Figure 6a. After mapping out the atoms carefully, we figure out that it is corresponding to one of the structures with 1/8 Fe-vacancy state, i.e. the $\sqrt{8}\times\sqrt{10}$ parallelogram block (in units of the Fe-Fe bonds). In Figure 6b, we plot the K atoms on top of the Fe vacancies, together with the Se atoms between the K and the Fe layers. As shown in Figure 6a, although there are some domain walls in the field of view, this kind of structure has been actually observed in most places of this region. Although our STM data here cannot give a direct evidence for the 1/8 Fe-vacancy $\sqrt{8}\times\sqrt{10}$ state, the expectation of the composition of this state is very consistent with the chemical composition obtained from the superconducting domains, that is $K_{0.68}Fe_{1.78}Se_2$. Therefore we would conclude that there is a strong indication that the superconducting phase is derived from the parent phase $K_{0.5}Fe_{1.75}Se_2$ by partially filling K and Fe to the unoccupied sites, which induces doping of electrons into the system. The parent state $K_{0.5}Fe_{1.75}Se_2$ is naturally charge balanced and the K concentration here may be just high enough to sustain the 122 structure. Any extra but slight doping to the K-sites and the Fe-sites in $K_{0.5}Fe_{1.75}Se_2$ will induce electron doping, strengthen the stability of the 122 structure but with the tolerance of the charge imbalance to a certain extent. Superconductivity is achieved by the electron doping and under such a subtle balance. We must emphasize that, although we have observed a unique 1/8 Fe-vacancy $\sqrt{8}\times\sqrt{10}$ state with the parallelogram structure, however, we cannot rule out the existence of other possible structures completely. The observation of this very unique pattern in such a large area strongly suggests that this structure should be a very competent one with lower energy, which may be resolved by the future theoretical calculations. Furthermore, we need to clarify that the pattern of K atoms on the surface play here only as the indicators of the Fe vacancies in the beneath layer of Fe, one should not extend this structure for the K atoms in the body of the sample.

Concerning the electronic properties of the $K_{0.5}Fe_{1.75}Se_2$ state (or called as the 278 phase), at this moment, we can only have limited speculations. First of all, this phase may not have an antiferromagnetic transition at temperatures below 300 K, since we have not found any such transitions from our measurements. This is qualitatively consistent with the recent calculations on the 278 phase[30]. In addition, in this phase, compared with the 245 state, the Fe vacancies are much diluted, therefore one can expect that it may be more "metallic" and gapless, or may have a much smaller gap, but should still be a bad metal. Recently, it is shown that by intercalating the so-called neutral layers, like molecules $Li_x(NH_2)_y(NH_3)_{1-y}$, between the FeSe layers can produce the 122 structure with much higher Fe concentration without inducing large electrostatic force and thus lead to the superconducting samples with larger superconducting shielding volume[31,32,33]. This development is certainly consistent with our general arguments. It remains to know what basic properties should exhibit by the proposed 1/8 Fe-vacancy $\sqrt{8}\times\sqrt{10}$ state. Our discovery here should stimulate further theoretical and experimental investigations.

## Methods

**Sample synthesis**. The original $K_xFe_{2-y}Se_2$ single crystals were fabricated by using the self-flux method. Firstly, FeSe was prepared as the precursor by reacting Fe powders (purity 99.99 %) and Se grains (purity 99.99 %) in the ratio of 1:1 at 700 °C for 24 hours. Then the starting materials with an atomic ratio of K:FeSe = 0.8:2 were loaded into an alumina crucible and sealed in a quartz tube under vacuum. Sometimes a second bigger quartz tube is used in case the inner quartz tube cracks during the sintering. All the weighing, mixing, grinding and pressing procedures were finished in a glove box under argon atmosphere with the oxygen and moisture below 0.1 PPM. The mixture were subsequently heated up to 1030°C and held for 3 hours. The sample was cooled down to 800 °C at a rate of 4 °C/h, and then it was cooled to room temperature by switching off the power of the furnace.

**Structure characterization.** A Zeiss Ultra-55 field-emission scanning electron microscopy (FE-SEM) was employed to measure the microstructures of the samples. With the 10 kV accelerating voltage and an AsB detector working in the topographic mode, domain structures with high contrast were observed. An Oxford Instruments Inca X-Max 50 energy-dispersive X-ray spectroscope (EDS) installed on the SEM was used to evaluate lateral distributions of the elements in the samples. The EDS measurements were performed with the spot and linear modes, respectively. In the spot mode, 50 spots were randomly selected on and off the domains of each sample to obtain high resolution spectra with a measurement time of 1 min for one point, with the voltage of 20 kV. In the line mode, the scanning was performed on lines across randomly selected domains with a dwell time of 10 s per step.

**STM measurements.** The scanning tunneling microscopy experiments were carried out with an ultrahigh vacuum, low temperature and high magnetic field scanning probe microscope USM-1300 (Unisoku Co., Ltd.). All processes before transferring the sample into the load-lock of the STM for cleaving were carried out in a glove box with Ar gas for protection. Then the samples were loaded to the STM system and cleaved at room-temperature in an ultra-high vacuum with a base pressure of about $1\times10^{-10}$ torr. After cleaving, the sample was immediately transferred to the STM measuring stage at the bottom of the cryostat. In the STM measurements, Pt/Ir tips were used.

**Acknowledgements**

We acknowledge the useful discussions with Igor Mazin, Fuchun Zhang, Zhongyi Lu, Tao Xiang, Hong Ding and Jiangping Hu. We appreciate the kind help from Mu Wang for the help in using the FE-SEM. This work is supported by NSF of China (A0402/11034011, A0402/11190023), the Ministry of Science and Technology of China (973 projects: 2011CBA00100, 2012CB821403, 2012CB21400) and PAPD.

**Author contributions**

The sample preparation and the follow-up thermal treatment were done by XXD. The transport,

magnetic measurements were done by XXD. The SEM measurements and analysis were carried out by XXD, GBM, CM and YHH. The low-temperature STS measurements were finished by DLF, ZYW and HY. HHW designed and partially joined the experimental procedures, coordinated the whole work, wrote the manuscript. All authors have discussed the results and the interpretation.

## Competing financial interests

The authors declare that they have no competing financial interests.

## Figure Legends

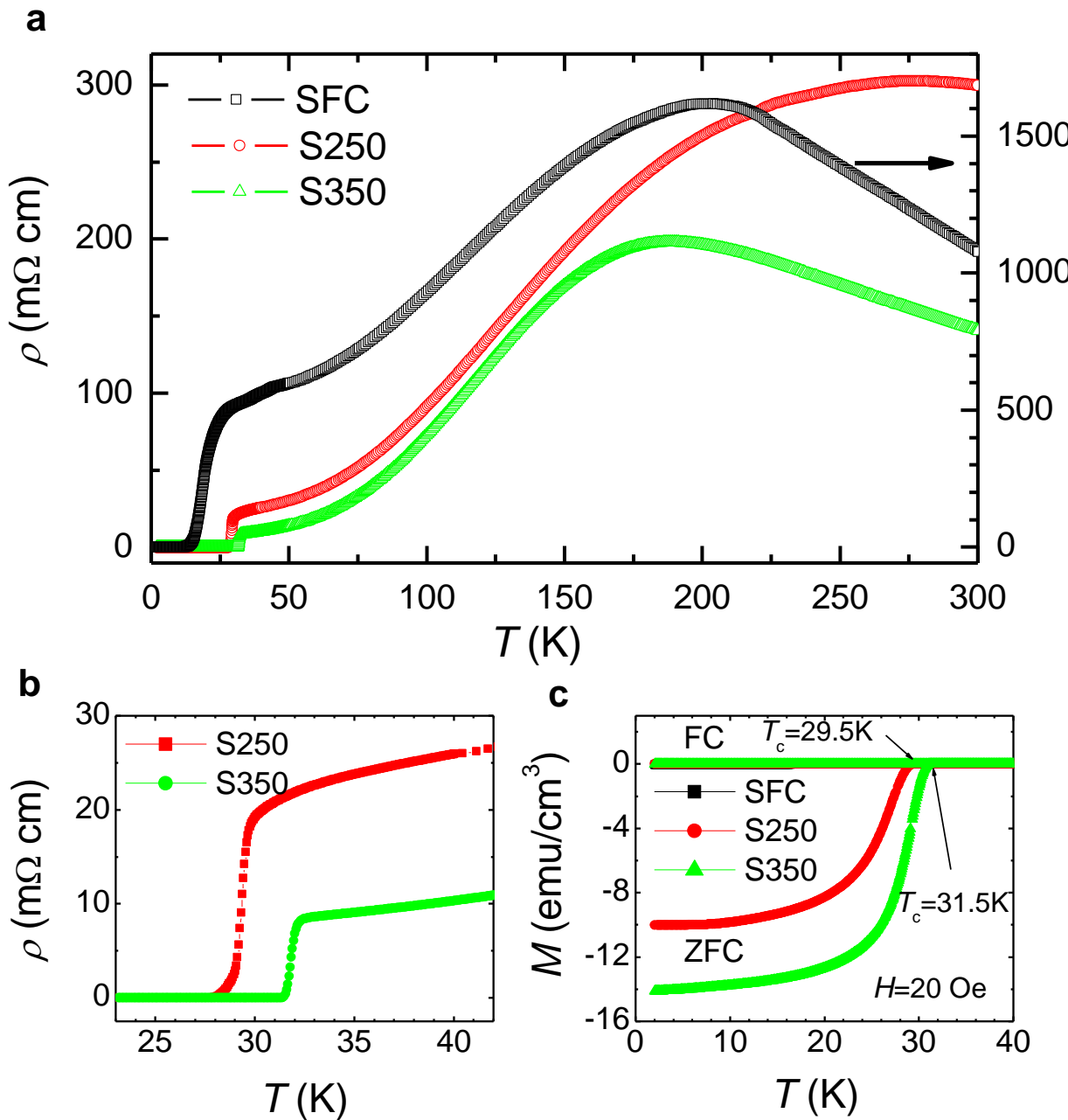

**Figure 1 │ Resistive and magnetic properties of the three typical samples. a,** Temperature dependence of resistivity of the three typical samples SFC, S250 and S350 experiencing different thermal treatments. The slow furnace cooled sample SFC has a rather broad transition width and a large residual resistivity. **b,** The resistive data shown in an enlarged view in low temperature region. In contrast, the sample S350 has a rather sharp superconducting transition at about 32 K with a small residual resistivity. **c,** Temperature dependence of DC magnetization of the three samples measured with the zero-field-cooling (ZFC) and field-cooling (FC) modes at $H$ = 20 Oe. Since the ZFC signal of the SFC sample is very small, it overlaps with the FC curves of the three samples.

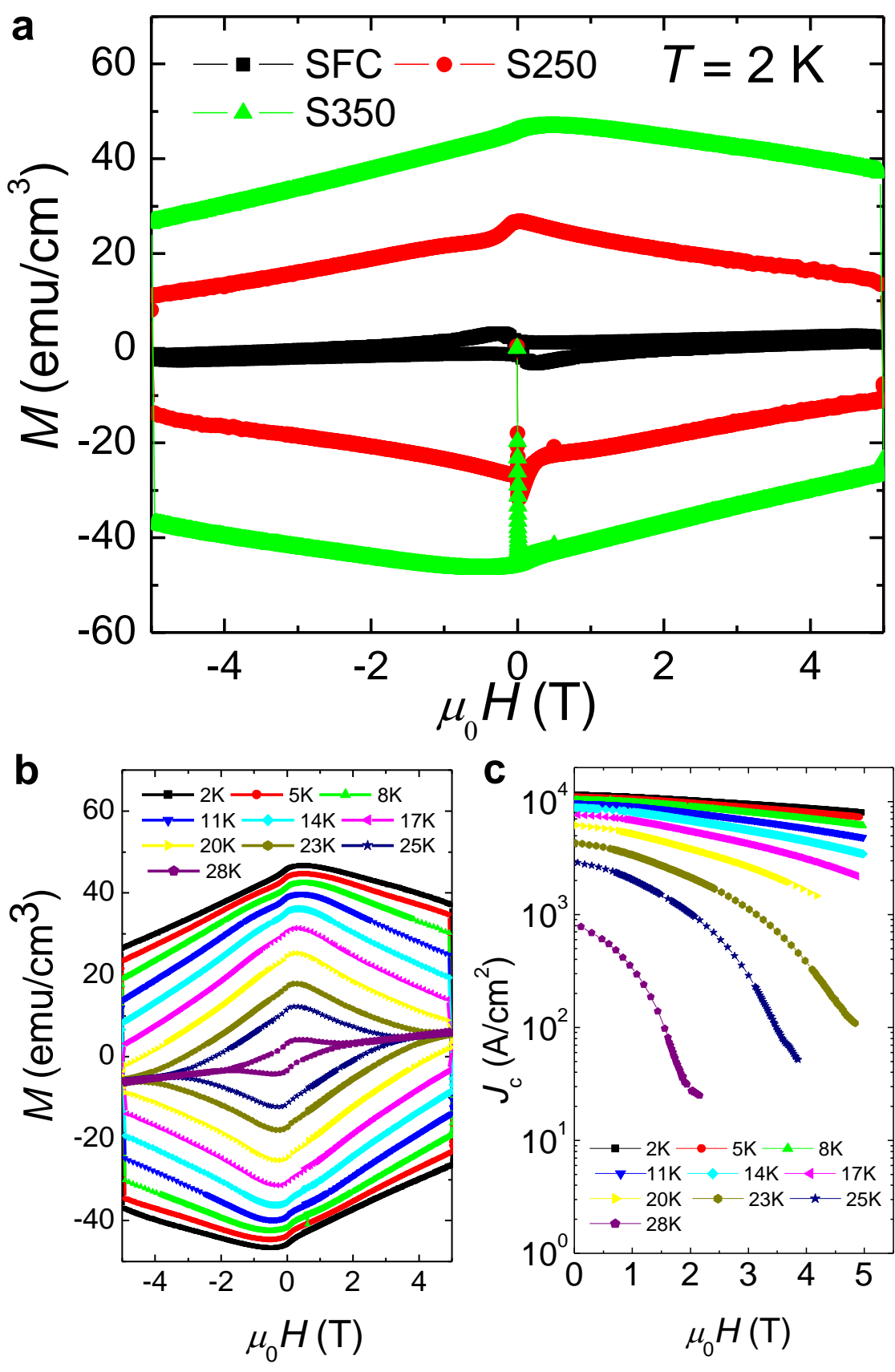

**Figure 2 | Magnetizations of the three typical samples.** (**a**) The magnetic field dependence of the magnetization of the three samples SFC (square symbols), S250 (diamond symbols) and S350 (triangle symbols). (**b**) The magnetization-hysteresis-loops of the sample S350 measured at temperatures from 2 K to 28 K. A magnetization hysteresis can be observed up to 2 T even at 28 K, being close $T_c$. (**c**) The magnetic field dependence of the critical current density calculated based on the Bean critical state model, although it may be invalid for this sample with phase separation. The purpose is to give a rough comparison between the present sample S350 and that of the FeAs-based 122 samples.

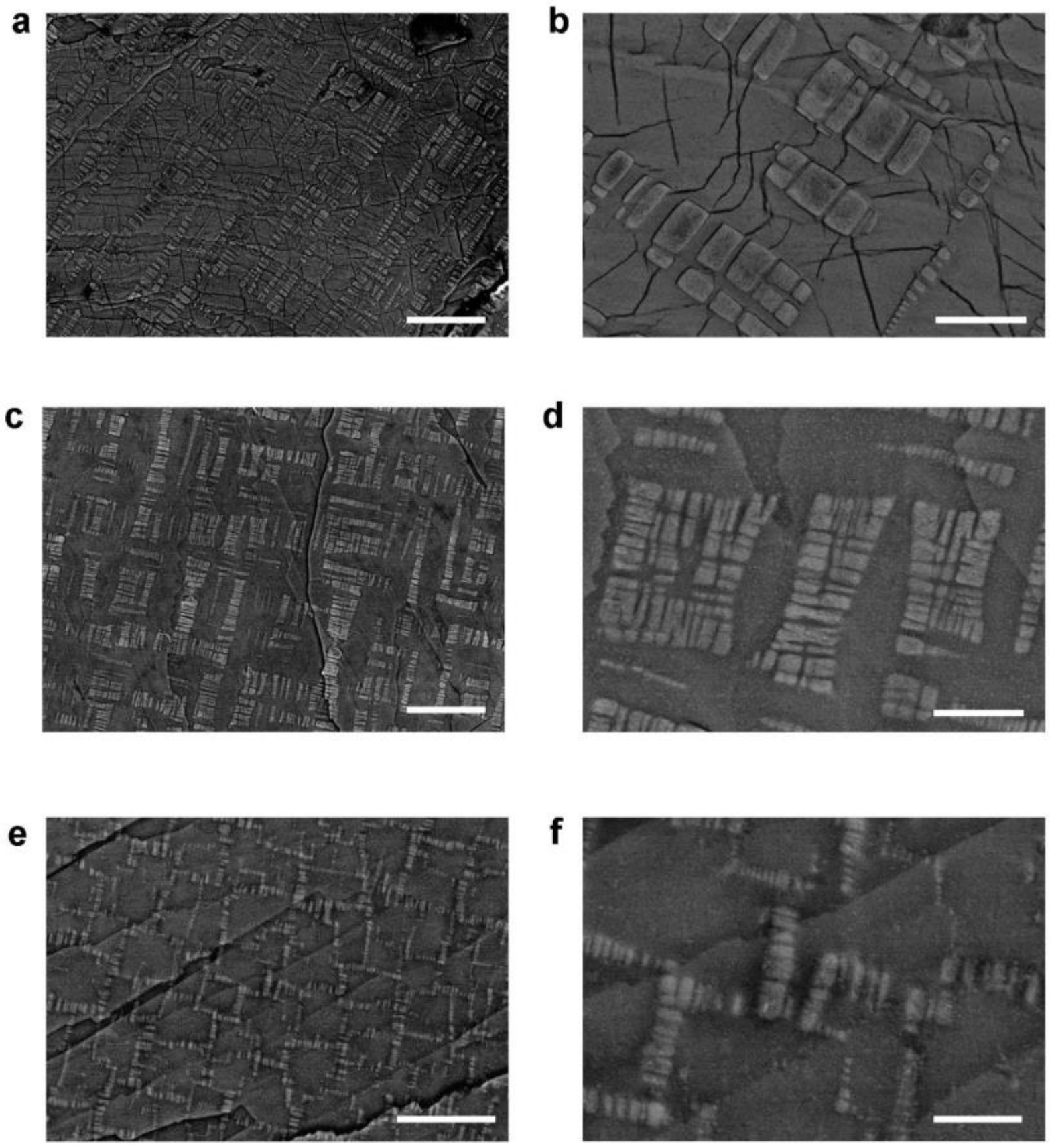

**Figure 3 │ Back-scattered electron images of SEM measurements on the cleaved surface of three typical samples.** (**a, b**) The topography of the cleaved surface of the sample SFC. It is clear that the sample separates into two kinds of regions: the domains with brighter color and rectangular shape, which are buried in the background of a darker region. Some cracks can be found in the background of the sample. The domains are widely separated. Scale bar, (**a**) 10 μm, (**b**) 2 μm. (**c, d**) The topography of the cleaved surface of the sample S250. The domains with brighter color in the sample of SFC further crack into many smaller domains, but are still far apart from each other. Scale bar, (**c**) 10 μm, (**d**) 2 μm. (**e, f**) The SEM image of the sample S350. One can see that the domains become very small and well connected and form some networks which look like the spider-web. Scale bar, (**e**) 5 μm, (**f**) 1 μm. The measurements were done with the voltage of 10 kV**.**

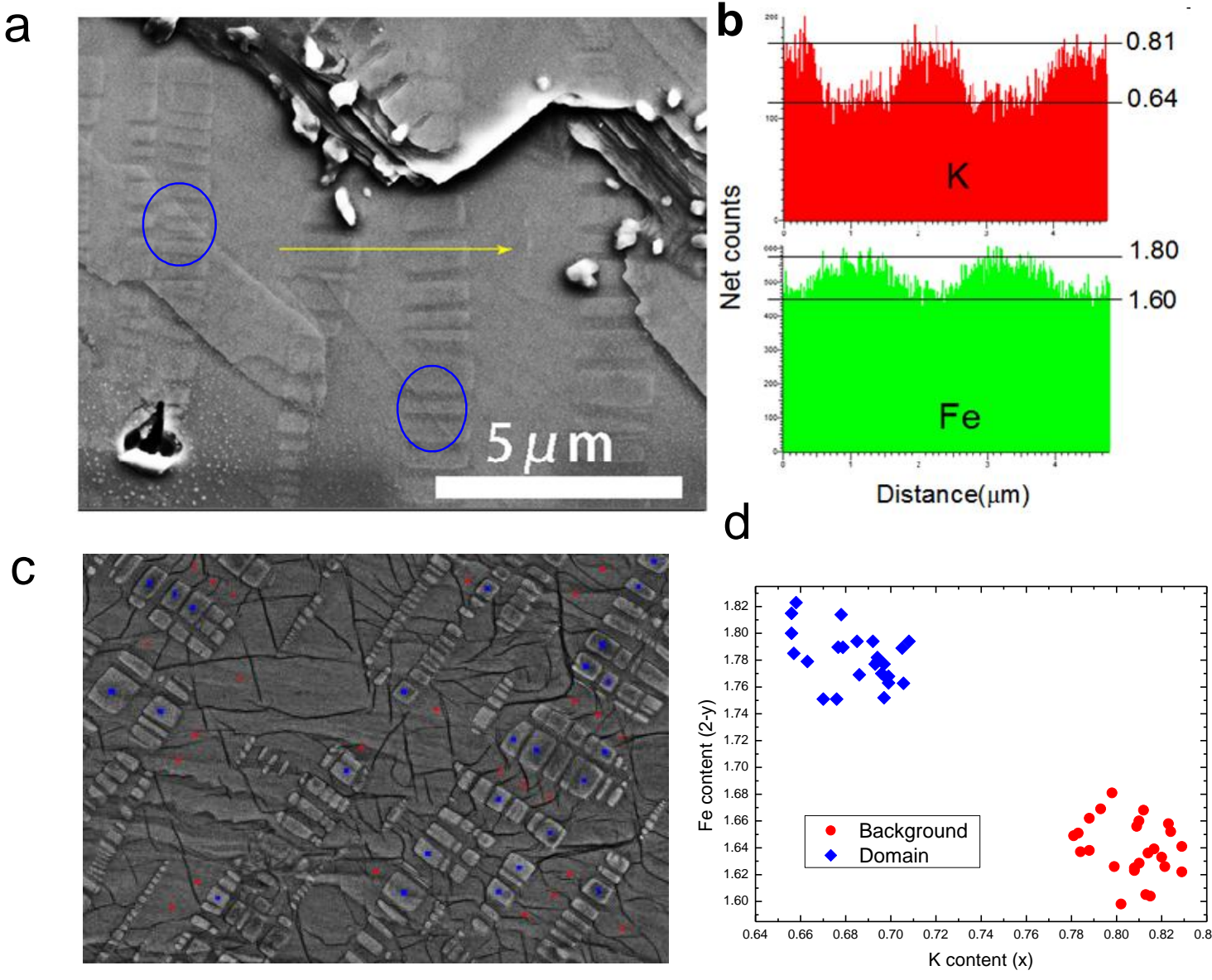

**Figure 4 │ Correlations between the microstructure and the analysis on the compositions.** (**a**) The topography of one cleaved surface of a sample SFC. Scale bar, 2 μm One can clearly see the rectangular domains that are buried in the background. The yellow arrowed line highlights the trace along which the spatial distribution of compositions of K and Fe are measured and presented in (**b**)**.** The large blue circles here mark the positions where the rectangular domains go through several layers along c-axis. The outline and shade of the domains can be seen on the layers with different height. (**c**) The SEM image of the sample SFC in another region. The red spots and blue spots mark the positions where the local compositions are analyzed. (**d**) The compositions of K and Fe measured on the rectangular domains (blue diamond) and the background (red circles). One can clearly see that the data fall into two groups: the formula $K_{0.68}Fe_{1.78}Se_2$ on the domain, and $K_{0.81}Fe_{1.8}Se_2$ on the background. The measurements are done with the voltage of 20 kV.

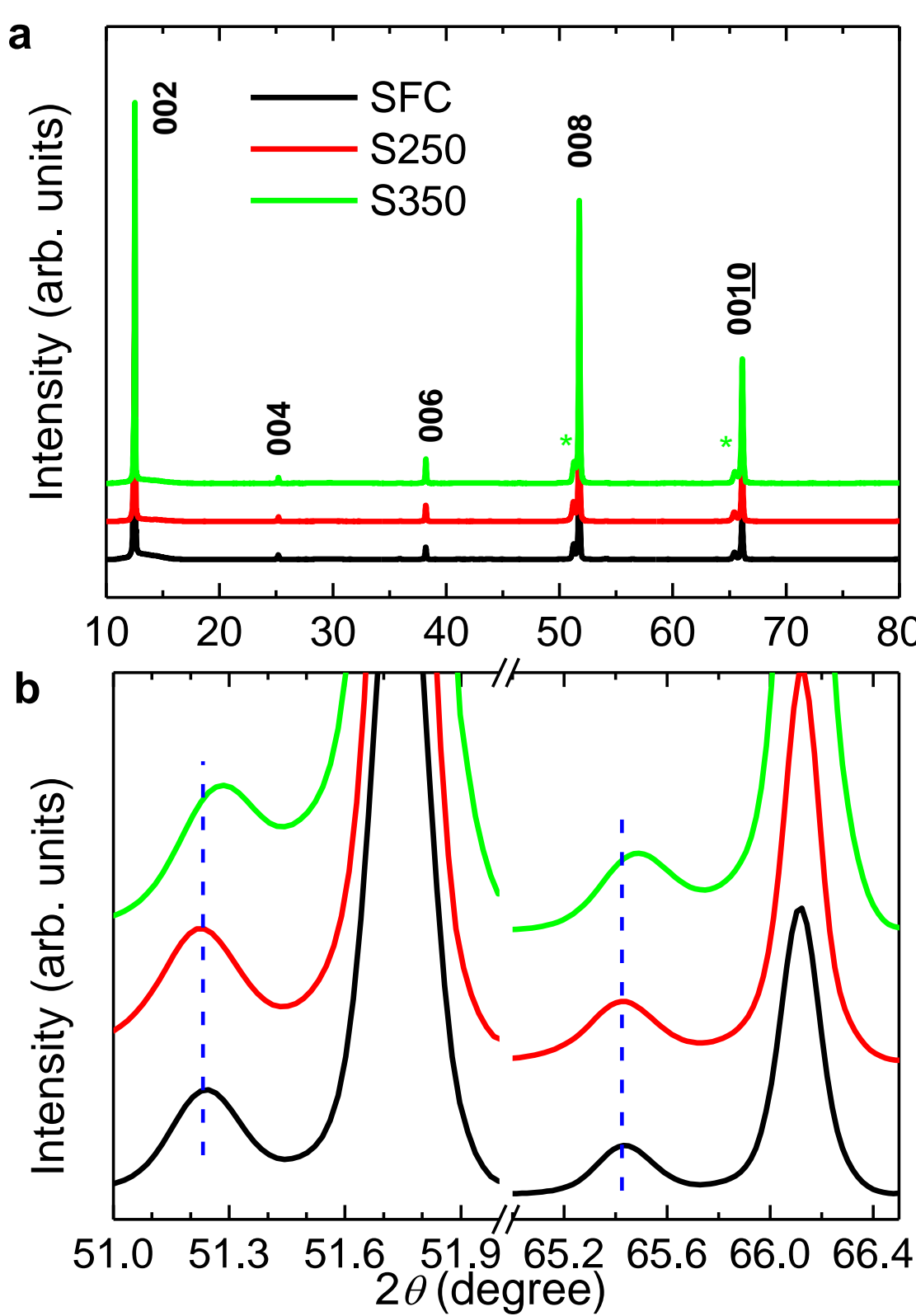

**Figure 5│ X-ray diffraction patterns of the $K_xFe_{2-y}Se_2$ single crystals.** (**a**) the (00*l*) reflections from the basal plane of the samples SFC, S250 and S350, respectively. The asterisks indicate the peaks of the superconducting phase. (**b**) An enlarged view of the (008) and ( $00\underline{10}$ ) reflections.

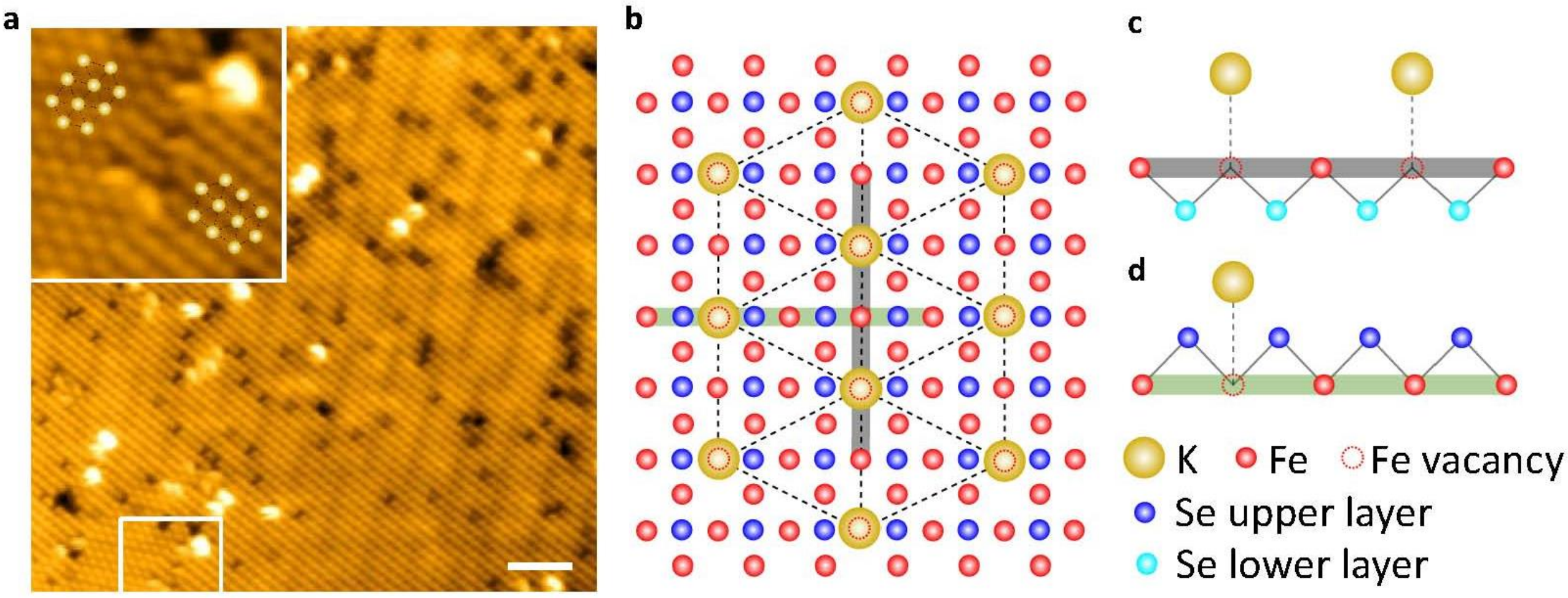

**Figure 6 | Atomically resolved topography and the sketch of the 1/8 Fe-vacancy $\sqrt{8}\times\sqrt{10}$ structure.** (**a**) The measured STM image on the [001] surface of the sample SFC with the proposed 1/8 Fe-vacancy $\sqrt{8}\times\sqrt{10}$ structure. The bias voltage and tunneling current during the measurements are fixed at 4.2 V and 50 pA, respectively. The image on the upper-left corner is an enlarged view of the region enclosed by the frame on the bottom. In the inset we use the white circles to represent the atoms of K. On the surface there are mainly two domains with different orientations, but they have the same structure, that is the $\sqrt{8}\times\sqrt{10}$ parallelogram block. Scale bar, 5 nm. (**b**) Sketch of the K, Fe and Se atoms with the Fe vacancies in the beneath layer. The big yellow circles and hollow red circles with dashed-outlines represent the K atoms and the Fe vacancies respectively. In our proposal they have the same $\sqrt{8}\times\sqrt{10}$ parallelogram structure. (**c, d**) The partial structure constructed by the K, Fe and Se atoms when K resides just above the Fe vacancy, by crossing the two cuts along the traces highlighted by the thick grey and light-green lines in (**b**).

## Supplementary Materials

**Author correction: Influence of microstructure on superconductivity in $K_xFe_{2-y}Se_2$ and evidence for a new parent phase $K_2Fe_7Se_8$**

Xiaxin Ding, Delong Fang, Zhenyu Wang, Huan Yang, Jianzhong Liu, Qiang Deng, Guobin Ma, Chong Meng, Yuhui Hu & Hai-Hu Wen
*Nature Communications* **4,** 1897 (2013).

In light of concerns that have been raised regarding the measurement, analysis and presentation of the sample compositions from energy-dispersive X-ray spectra (EDS) in Fig.4, the authors would like to add a number of clarifications and amendments to the published version of this Article:

In the original version of this Article, the thirteenth, fourteenth, fifteenth and sixteenth sentences of the first paragraph of the 'Scanning electron microscope measurements' section of the Results currently read 'As presented in Fig. 4a,b, close correlation between the local compositions of K and Fe is observed. By figuring out the local compositions, we find that the brighter domains have a composition of about $K_{0.64}Fe_{1.78}Se_2$, while the background area has a composition of about $K_{0.81}Fe_{1.60}Se_2$, the latter is very close to the standard $K_2Fe_4Se_5$ phase. To strengthen these conclusions, we did the local analysis on 50 randomly selected specific points, marked by the red spots (on background) and blue spots (on the domains) in Fig. 4c. The statistics on these data are given in Fig. 4d. Interestingly, the data are rather converged and fall into mainly two groups, one with the composition of about $K_{0.68}Fe_{1.78}Se_2$ and another one around $K_{0.8}Fe_{1.63}Se_2$.'

We apologize that these sentences are imprecise. The quantitative numbers in Fig. 4b cannot be obtained from the line scan profile data. Therefore, the four horizontal lines in Fig. 4b should be ignored. We have also added to this panel the spatial distribution of Se composition along the same line. We have also revised Fig. 4d to use all 79 data points to analyze the compositions. Therefore, the averaged compositions after the EDS analysis has been corrected from $K_{0.68}Fe_{1.78}Se_2$ to $K_{0.69}Fe_{1.78}Se_2$ for domains, and from $K_{0.80}Fe_{1.63}Se_2$ to $K_{0.80}Fe_{1.65}Se_2$ in background areas.

The correct version of these sentences should read 'As presented in Fig. 4a,b, close correlation between the local compositions of K and Fe is observed; that of Se is relatively more uniform. Although we did the line scan measurements of compositions of K and Fe following the yellow arrowed line in Fig.4a, it is known that the line scan measurements usually cannot tell precise compositions. To get more accurate compositions, we did the local analysis on 79 selected points, among them 39 points are on the domains and 40 in the background region. The statistics on these data are given in Fig. 4d. Interestingly, the data are rather converged and fall into mainly two groups, one with the average composition of about $K_{0.69}Fe_{1.78}Se_2$ and another one around $K_{0.8}Fe_{1.65}Se_2$.'.

The last sentence of this same paragraph reads 'Further analysis on the sample S350 gives the similar compositions in two different regions as the SFC.' We apologize for the imprecise description of this in the paper. We did not include any EDS data for sample S350 in Fig.4. The correct version of this sentence should read 'Further analysis based on grey-scale and XRD measurements suggest that the sample S350 may have similar compositions in the two different regions as the SFC, which is detailed below.'

In light of the above errors, the corrected version of Fig. 4 and its caption are:

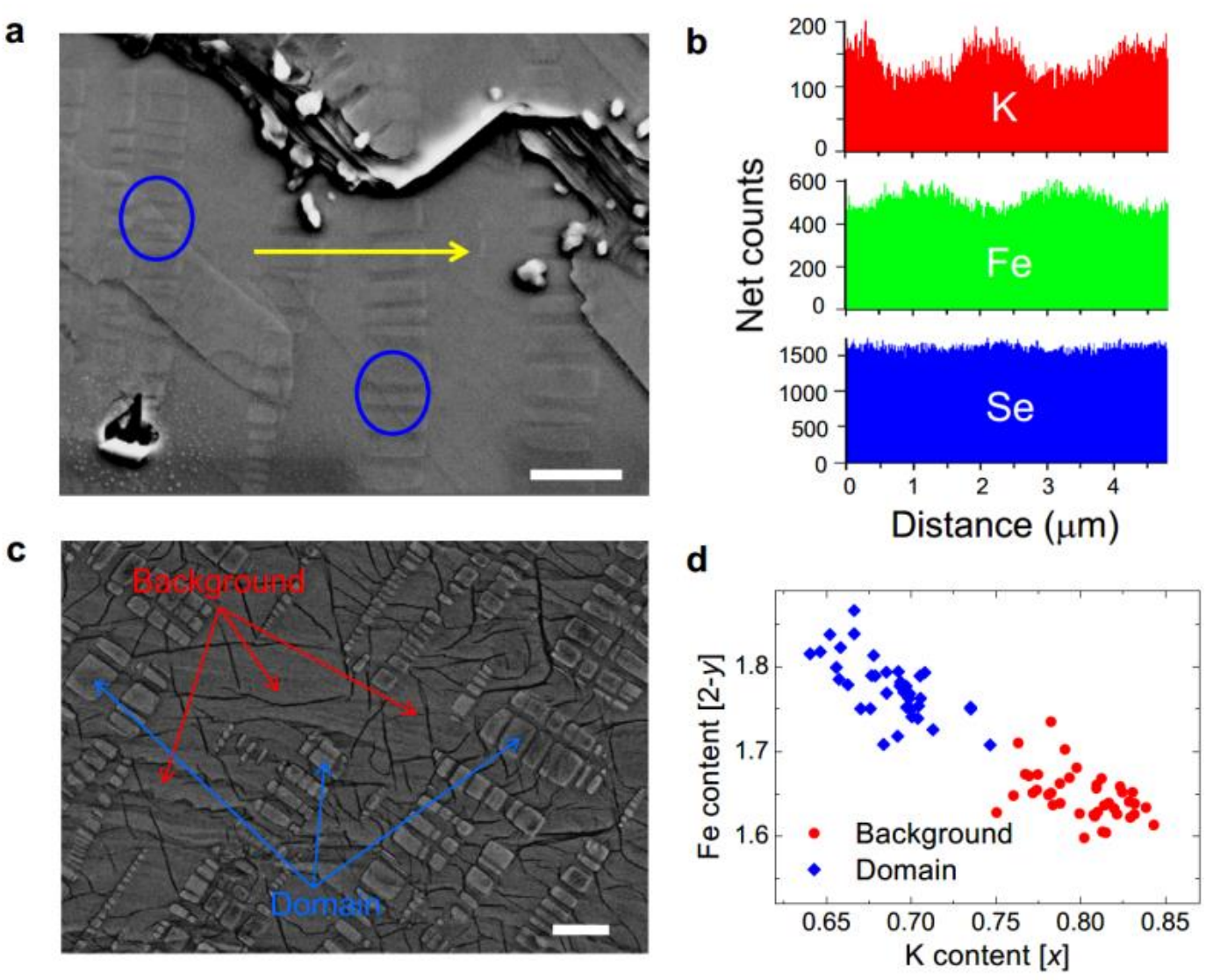

Figure 4 | Correlations between the microstructure and the analysis on the compositions. (a) The topography of one cleaved surface of a sample SFC. Scale bar, 2 μm. One can clearly see the rectangular domains that are buried in the background. The yellow arrowed line highlights the trace along which the line scan profile of the compositions was taken, as shown in (b). The large blue circles here mark the positions where the rectangular domains go through several layers along c axis. The outline and shade of the domains can be seen on the layers with different height. (c) Representative SEM image (taken with an accelerating voltage of 10 kV) of the typical domain (marked by blue arrows) and background (marked by red arrows) regions. From these two different regions the EDS spectra were collected for point-wise composition analysis. Scale bar, 2 μm. (d) The compositions of K and Fe were measured from the total 79 spots on the domain and on the background regions of three SFC samples cleaved from a single bulk. One can clearly see that the EDS data fall into two groups. By analyzing the total EDS dataset (39 points on the domains and 40 points in the background region), we get the averaged compositions with the formula $K_{0.69}Fe_{1.78}Se_2$ on the domain, and $K_{0.80}Fe_{1.65}Se_2$ on the background, respectively. The measurements are done with the voltage of 25 kV.

Specifically, the fourth sentence in the original legend of Fig. 4 currently reads 'The yellow arrowed line highlights the trace along which the spatial distribution of compositions of K and Fe are measured and presented in **(b)**.' In the corrected version of the legend, this sentence reads 'The yellow arrowed line highlights the trace along which the line scan profile of the compositions was taken, as shown in (b).'

The seventh and eighth sentences of the original legend read '**(c)** The SEM image of the sample SFC in another region. The red spots and blue spots mark the positions where the local compositions are analysed.' This is incorrect: these 25 red spots and 25 blue spots do not represent the precise locations from which EDS data were collected. In the corrected version of the legend, these two sentences state '**(c)** Representative SEM image (taken with an accelerating voltage of 10 kV) of the typical domain (marked by blue arrows) and background (marked by red arrows) regions. From these two different regions the EDS spectra were collected for point-wise composition analysis.'

The tenth and eleventh sentences of the original legend read '**(d)** The compositions of K and Fe measured on the rectangular domains (blue diamond) and the background (red circles). One can clearly see that the data fall into two groups: the formula $K_{0.68}Fe_{1.78}Se_2$ on the domain, and $K_{0.81}Fe_{1.6}Se_2$ on the background.' In the original version of Fig. 4d, we presented the composition obtained from 50 different EDS data (25 from domains and 25 from background regions) that were selected from a total of 79 point-wise EDS measurements (40 on the background and 39 on the domains) collected from three samples cleaved from a single bulk SFC sample, instead of one sample indicated in the original paper. We want to clarify that there is no position-wise correlation between these data and those points marked in Fig. 4c of the original paper. We apologize for not presenting the full set of data points in the paper, which is a clear deviation of standard practice. The corrected version of the legend replaces these two sentences with '**(d)** The compositions of K and Fe were measured from the total 79 spots on the domain and on the background regions of three SFC samples cleaved from a single bulk. One can clearly see that the EDS data fall into two groups. By analyzing the total EDS dataset (39 points on the domains and 40 points in the background region), we get the average compositions with the formula $K_{0.69}Fe_{1.78}Se_2$ on the domain, and $K_{0.80}Fe_{1.65}Se_2$ on the background, respectively.'

The final sentence of the original legend reads 'The measurements are done with the voltage of 20 kV.' The corrected version states '25 kV' instead of '20 kV'.

In addition to these errors in Figure 4 and the related discussion, the fourth sentence in the third paragraph of the 'Scanning electron microscope measurements' section of the Results currently reads: 'Similar behaviour of the composition distribution is observed in the sample S350, which presents better global appearance of superconductivity, although now the domains become much smaller.' This is incorrect. This sentence should be replaced with: 'Similar grey-scale fraction is observed in samples SFC and S350, which may suggest similar compositions of the superconducting areas of SFC and S350, the latter presents better global appearance of superconductivity, although now the domains become much smaller.'

The first sentence of the Discussion reads 'In our local analysis presented above, it is found that the composition of the superconducting domains are converged at about $K_{0.68}Fe_{1.78}Se_2$.' It should say '$K_{0.69}Fe_{1.78}Se_2$' instead of '$K_{0.68}Fe_{1.78}Se_2$'.

The fourth and fifth sentences of the Discussion read 'Concerning the uncertainties in these totally distinct techniques, the consistency between the two techniques are remarkable, which also validates the argument that the domains are superconducting and have a composition of about $K_{0.68}Fe_{1.78}Se_2$. In the background phase, the analysis shows a composition of $K_{0.8}Fe_{1.63}Se_2$, which is naturally attributed to the well-known 245 phase.' These should say '$K_{0.69}Fe_{1.78}Se_2$' instead of '$K_{0.68}Fe_{1.78}Se_2$' and '$K_{0.8}Fe_{1.65}Se_2$' in place of '$K_{0.8}Fe_{1.63}Se_2$'.

The seventeenth sentence of this same paragraph reads 'A glance at our formula $K_{0.68}Fe_{1.78}Se_2$ of the superconducting domains would suggest that a possible phase with the formula of $K_{0.5}Fe_{1.75}Se_2$ (or called as 278 phase) is next to the $K_{0.8}Fe_{1.6}Se_2$ (or the 245) phase and may act as the parent phase for superconductivity.' It should say '$K_{0.69}Fe_{1.78}Se_2$' instead of '$K_{0.68}Fe_{1.78}Se_2$'.

The twelfth sentence of the second paragraph in the Discussion reads 'Although our STM data here cannot give a direct evidence for the 1/8 Fe-vacancy $\sqrt{8}\times\sqrt{10}$ state, the expectation of the composition of this state is very consistent with the chemical composition obtained from the superconducting domains, that is, $K_{0.68}Fe_{1.78}Se_2$.' It should say '$K_{0.69}Fe_{1.78}Se_2$' instead of '$K_{0.68}Fe_{1.78}Se_2$'.

The fifth sentence of the 'Structure characterization' section of the Methods currently reads 'In the spot mode, 50 spots were randomly selected on and off the domains of each sample to obtain high resolution spectra with a measurement time of 1 min for one point, with the voltage of 20 kV.' This should state '79 spots were selected' instead of '50 spots were randomly selected', 'of three SFC samples' instead of 'of each sample' and '25 kV' instead of '20 kV'.

Finally, the third sentence of the Author Contributions section originally read 'The SEM measurements and analysis were carried out by X.D., G.M., C.M. and Y.H.' This should instead read 'The SEM and line scan measurements were carried out by X.D., G.M. and C.M.; X.D. and Y.H. did the analysis of the EDS data and made their connection to the SEM measurements.'

These errors have not been corrected in the PDF or HTML versions of the Article.